\documentstyle[prd,eqsecnum,aps]{revtex}
\begin{document}
\draft
\preprint{IUCAA 10/1994}
\title {Performance of Newtonian filters in detecting gravitational \\
 waves from coalescing binaries}
\author{R. Balasubramanian and S. V. Dhurandhar}
\address{Inter-University Centre for Astronomy and Astrophysics, \\
Post Bag 4, Ganeshkhind, Pune 411 007, India}
\date{\today}
\maketitle
\begin{abstract}
As coalescing binary systems are one of the most promising sources of
gravitational waves, it becomes necessary to device efficient detection
strategies. The detection strategy should be efficient enough so as not
to miss out any detectable signal and at the same time minimize the false
alarm probability. The technique of matched filtering used in the detection
of gravitational waves from coalescing binaries relies on the construction of
accurate templates. Until recently filters modelled on the quadrupole or the
Newtonian  approximation were deemed sufficient. Such filters or templates have
in addition to the amplitude, three parameters which are the chirp mass, the
time of arrival and the initial phase. Recently it was shown that
post-Newtonian
effects contribute to a secular growth in the phase difference
between the actual signal and its corresponding Newtonian template.
This affects the very foundation of the technique of matched filtering,
which relies on the correlation of the signal with the filter and hence is
extremely sensitive to errors in phase. In this paper we investigate the
possibility of compensating for the phase difference
caused by the post-Newtonian terms by  allowing for a shift in the Newtonian
filter parameters. The analysis is carried out for cases where one of the
components is a black hole and the other a neutron star or a small black
hole. The alternative strategy would be to increase the
number of parameters of the lattice of filters which might prove to be
prohibitive in terms of computing power.  We find that Newtonian  filters
perform adequately  for the purpose of detecting the presence of the signal
 for both the initial and the advanced LIGO detectors.
\end{abstract}
\section{Introduction}
Coalescing binaries are the most promising sources of gravitational
waves \cite{1} for laser interferometric gravitational wave detectors.
The basic reason for the importance of these type of sources is their
broadband nature which makes them  ideally suited for their detection by the
interferometers. The binary systems which are of relevance here are those
consisting of compact objects i.e. black holes and neutron stars. It has been
estimated that three such coalescences occur per year out to  a  distance of
200 Mpc  \cite{15,16}. A lot of  attention has
recently been focussed on the issues of
detecting the presence of the signal and the extraction of  astrophysical
information from the estimated parameters of the signal.

There are plans to construct such laser interferometers around the  globe and
by
the end of this century the American LIGO \cite{3} and French/Italian VIRGO
\cite{4} will be in operation. The emphasis in their construction is on the
reduction of noise which may be thermal, seismic,  quantum, or  photon shot
noise. In laser interferometric  detectors the
lower cuttoff is decided   by the seismic noise which is very dominant
at low fequencies. It is expected that the LIGO will be able to go down
to 40 Hz in its initial stage and to 10 Hz  in its final stage. This means
that that we can start observing the binary when its orbital frequency
is 20 Hz in the case of the initial detectors and 5 Hz in the case
of the advanced ones. This leads to sufficiently large integration times
which enhances the signal to noise ratio. It was suggested
by Thorne that matched filtering would be an ideal filtering technique
for this purpose.   Matched filtering is a standard technique used in signal
analysis when the waveform is known. It determines for us an optimal linear
filter which can decide on the presence or absence of the signal waveform in a
given data train \cite{7,8,11,12,20}. This requires accurate
modelling of the waveforms, which is possible for the coalescing
binary systems. They are clean systems and their inspiral waveform depends
on a few parameters such as the individual masses and spins. Tidal interactions
do not matter until the very end of the inspiral \cite{21,24}.
A lot of research activity has gone in the direction
of obtaining accurate templates under the various approximation schemes
such as the quadrupole and the post-Newtonian \cite{25,2,9,10}.
Recently it has been shown that post-Newtonian (PN) corrections, spin-orbit
(S.O.) and spin-spin (S.S.) couplings, produce in the
waveform an accumulating phase error as compared to the
Newtonian expression \cite{2}. Therefore, a template constructed from the
Newtonian waveform would go out of phase with the signal and the
so called ``matched filtering'' technique for detection would
woefully fail. In this paper we show that as long as we are only
{\em searching} for signals a Newtonian filter would perform remarkably
well even though the signal contains PN corrections. The key idea here
is that we allow the parameters of the Newtonian filter to vary and adjust
so as to produce the maximum possible correlation with the signal.
We have found that this flexibility allows for fairly high values of the
correlation. In many cases of interest the correlation obtained is  70\% of
its maximum possible value which would have been obtained had the
template been  perfectly matched to the signal. On the other hand,
a template with the same parameters as those of the signal produces
correlations of about 10 to 20\%. We have carried out the analysis for
the two noise curves assuming a LIGO type detector. The two noise curves
are the power spectral densities of the noise for the LIGO in its initial
and advanced stages, as given in \cite{13}.
In the case of the initial LIGO detector the analysis is also carried out for
the case of white noise for the sake of comparision. Also a correspondence
between the parameters of the filter and the signal
could be set up, it might be possible to estimate the parameters of the signal
from those of the filter. In other words the filter parameters may be
``renormalized''.

	The paper is divided as follows. In section \ref {match}
 we elaborate on the chirp waveform, and the conventional detection strategy.
 We discuss the technique of matched filtering and define a quantity which
shall be a measure of how well a Newtonian waveform can match with a
post-Newtonian signal.We also make some comments about the signal
power spectrum. In section \ref {numeric} we discuss
the numerical results of the simulations carried out.
And finally in section \ref{end} we  summarise our results and indicate
future directions.

\section{Searching for the signal}
\label{match}
\subsection{Newtonian waveform and conventional stategy}
\label{conv}
	The waveform of the signal from the coalescing binary system henceforth
called the `chirp' has been modelled under various approximations.
In the quadrupole approximation the chirp has three parameters other than
the amplitude. These are  the initial phase $\phi_0$, the time of arrival $t_a$
 (i.e the time at which the instantaneous frequency of the equals the lower
cuttoff of the detector) and the coalescence time $\xi$ which form a
convenient
set of parameters for our purpose \cite{5,14}. The Newtonian waveform
$h(t;\xi,\phi_0,t_a)$ is given by,
\begin{equation}
h(t;\xi,\phi_0,t_a)={\cal N}a(t)^{-1/4}
cos[\frac{16\pi}{5}f_a\xi[1-a(t)^{5/8}] + \phi_0],
\end{equation}
where
\begin{equation}
\label{eqxi}
\xi = 3.003\left(\frac{{\cal M}}{M_\odot}\right)^{-5/3}\left(\frac{f_a}
{100 {\rm Hz}}\right)^{-8/3}{\rm sec},
\end{equation}
 and
\begin{equation}
a(t) = 1 - \frac{t - t_a}{\xi}.
\end{equation}
Here the lower cuttoff frequency is denoted by $f_a$, and ${\cal M} =
M^{2/5}\mu^{3/5}$  is called the chirp mass where $M$ is the total mass
and $\mu$ the reduced mass of the binary system. $M_\odot$ is the solar
mass and it is convenient unit for our purpose.

	Given the form of the signal and the statistical description of the
noise one has to design an adequate set of filters to detect the signal.
 The noise is assumed to be stationary and is further  specified by its power
spectral density $S_h(f)$ which is defined by the relation,
\begin{equation}
\overline{\tilde{n}(f)\tilde{n}^{*}(f^{'})}	=
S_h(f)\delta(f - f^{'}),
\end{equation}
where $\tilde{n}(f)$ is the Fourier transform of a particular
realization of noise and
the overbar indicates an ensemble average. The $S_h(f)$ defined
above is the two sided power spectral density. We are primarily in search of
a filter with an impulse response $q(t)$ which correlates best with the signal
i.e.
when the correlation as defined below takes its maximum value for a particular
value of time shift $\Delta t$.
\begin{equation}
	C(\Delta t) = \int_{-\infty}^{+\infty}h(t)q(t + \Delta t)dt.
\end{equation}
 This implies that the Fourier transform of the matched filter $q(t)$ to detect
the signal $h(t)$ is given by the relation,
\begin{equation}
	\tilde{q}(f) = \frac{\tilde{h}(f)}{S_h(f)} e^{2\pi if\Delta t}.
\end{equation}
For the numerical computations that follow we use the Fast Fourier
transform algorithm as given in {\it Numerical Recipes} \cite{22}. The
definition of  the Fourier transform is the same as given there
i.e.
\begin{equation}
\tilde n(f) = \int_{-\infty}^{+\infty}n(t)e^{2\pi ift}dt.
\end{equation}
The impulse response of the filter $q(t)$  depends on the
parameters $\xi$, $t_a$, $\phi_0$. It also depends the time shift $\Delta t$.
It now becomes important to judiciously space out the filters in the
parameter space keeping in mind the constraints of computing power.
Such an  analysis has been carried
 out in great  detail for both white and coloured noise by
Sathyaprakash and Dhurandhar (see \cite{5,14}). We
discuss briefly their major results:
\begin{enumerate}
\item It was found that $\xi$ the coalescence
time is a convenient  parameter to use since the filters are
equally spaced in this parameter, where the spacing is decided by a fixed drop
in the correlation.
\item For the  phase  $\phi_0$
we require just two filters for every value of the $\xi$ parameter, one
with $\phi_0 = 0$ and the other with $\phi_0 = \pi/2$. Due to their
orthogonality the correlation is maximised over the phase by simply taking
the square root of the sum of squares of the individual correlations, i.e.
\begin{equation}
	C(\Delta t) = \sqrt{C_0^2(\Delta t) + C_{\pi/2}^2(\Delta t)} \  ,
\end{equation}
where $C_0$ and $C_{\pi/2}$ are the correlations corresponding to filters with
phases $\phi_0 = 0$ and $\phi_0 = \pi/2$ respectively.
\end{enumerate}

We assume $t_a = 0$ in the design of the filter and
therefore the value of $\Delta t$ for which the maximum of the correlation
occurs is equal to the time of arrival of the signal. Such a procedure of
maximising the correlation over the phase and the time is carried out for each
value of $\xi$. The final maximization
of the correlation is then carried over the $\xi$ parameter. The set of
parameters for which the correlation is maximum are then presumed to be the
most likely values of the parameters of the gravitational wave signal.

\subsection{Post-Newtonian signal}
The post-Newtonian corrections lead to corrections to the phase and the
amplitude of the Newtonian signal and also lead to additive terms which
are qualitatively different from the quadrupole term. In the case of a general
binary system it is tedious and difficult to get the various corrections
to the evolution of the orbits of the binary. If one of the bodies
is large compared to the other as in the case of a black hole neutron star
binary system one can apply the Regge Wheeler
perturbation formalism \cite{23} to get the  evolution of the orbit. This
provides us with the evolution of the orbital frequency as a function of time.
This has been worked out \cite {2} and is given by
\begin{equation}
\label{orbev}
\frac{\dot f}{f^2} = \frac{96\pi}{5}\frac{\mu}{M}\frac{x^{2.5}}{F(x)},
\end{equation}
where,
\begin{equation}
\label {cut}
 F(x) = \frac{1 - \frac{3}{2}x - \frac{81}{8}x^2 - \frac{675}
{16}x^3}{1 - \frac{1247}{336}x + 4\pi x^{1.5} - 4.9x^2 - 38x^{2.5} + 135x^3}.
\end{equation}
Here $\dot f$ represents the first time derivative of frequency and
$x = (\pi Mf)^{2/3}$ the PN expansion parameter. The Newtonian waveform
is obtained from the above equation by setting $F(x) = 1$. The phase is
obtained by integrating equation (\ref{orbev}). For the amplitude we use the
Newtonian dependence on
the frequency {\it i.e.} $A(f(t) \approx$ const$\times  f^{2/3}$. This waveform
shall be called `restricted post-Newtonian' henceforth.   Although
 this is not exact, we do not expect the errors in the amplitude to
 affect the correlation significantly. We assume the initial phase and
the arrival time of the signal to be 0. As the matched filtering  process
can also be viewed as a correlation between the incoming signal and the
 filter it is evident that any secular growth of  phase difference will
reduce the correlation drastically.  Thus to have a matched filter
one must add one or more parameters. This would increase the number of filters
enormously  with  corresponding increase in computational time.
 It is worthwhile to explore whether we can substantially
increase the correlation by allowing for a shift in the
parameters of the
Newtonian filter i.e. whether the signal is able to achieve better
correlation with a Newtonian filter whose parameters are different from those
of
the signal. Obtaining large correlations depends on the function space
spanned by the signal and filter waveforms and to what extent they overlap.
 We obtain reasonably large correlations.
 Here  effects due to S.O. and S.S. coupling are
 not taken into account.  The addition of such terms will not alter the
thrust of the argument in that, some other Newtonian filter would  perform
 best.

\subsection{Filtering the post-Newtonian signal}
	A detailed account of the formalism and notation used here and the
theory of hypothesis testing using maximum likelihood methods as applied to
detection of gravitational waves from coalescing binaries is  given
in \cite{13,17}.
	We define a scalar product and its corresponding norm in the function
 space between two functions
$s(t)$ and $q(t)$ for future use;
\begin {equation}
\langle s(t),h(t)\rangle = \int_{-\infty}^{\infty}\frac
{\tilde s(f) \tilde h^{*}(f)}{S_h(f)}df,
\end {equation}
and
\begin {equation}
\|s\| = {\langle s(t),s(t)\rangle}^{1/2}.
\end {equation}
 If an exact matched filter
 were present then the signal to noise ratio $\rho$ would be simply equal
to $\|s\|$ where $s$ is the signal.
 Note  that our definition of SNR is different by a factor of two from the
one given in \cite{13} as they work with the one sided power spectral density.
The quantity we are interested in computing  is
\begin{equation}
\eta = \frac{\langle s,h\rangle} {\|s\| \|h\|},
\end{equation}
where $h(t)$ is the chirp corresponding to the filter. We shall term $\eta$
as the normalized correlation.
Henceforth when we use the word correlation we shall mean the quantity
$\eta$ unless specified otherwise.
As mentioned above the initial phase and the time of
arrival of the signal is taken to be 0. The aim is to maximise $\eta$ over
the range of parameters of the filter.
 The quantity $\eta$ takes the value between $0$ and $1$ and tells us
how well a Newtonian filter can substitute for a post-Newtonian one.
Geometrically one can visualize $\eta$ as the cosine of  the
angle between the signal vector and the chirp vector.

In figure \ref {fig1} we show the impulse response of a filter. As the noise is
 very high at lower frequencies the amplitude of the impulse response
 is very small at earlier times and becomes appreciable only at the end.
 Due to the same reason the increase of amplitude
with time is different from that of the Newtonian chirp.
Figure \ref{fig2} justifies the high correlations obtained. It shows
how well the filter matches with the restricted post-Newtonian signal.
It is to be noted that the filter matches with the signal very well
during the late stages where the amplitude is largest.

 Figure \ref{fig3} shows the power spectrum of the signal which is the square
of the magnitude of the Fourier transform of the signal divided by the power
 spectral density of noise as a function of frequency,
\begin {equation}
\Omega(f) = \frac{|\tilde s(f)|^2}{S_h(f)}.
\end{equation}
This quantity peaks near 200 Hz and it is in this frequency range therefore
that the filter must match the signal very well i.e. it should try to
keep in phase with the signal to yield a high correlation. This is
borne out in figure \ref{fig2}.
The stationary phase approximation used in the Fourier transform of the chirp
waveform \cite{5} predicts the power spectrum to be a smooth power law. This
however is not true, and numerical results and further investigations
into the stationary phase approximation bears this out. The stationary
phase analysis leads to a Fresnel integral and the smooth power law
fall off ($f^{-7/3}$) in the power spectrum is obtained if the limits of
integration extend from $-\infty$ to $\infty$. However since the chirp
waveform is taken to be of finite duration we actually get an incomplete
Fresnel
integral. This leads to oscillations in the power spectrum of the signal. The
oscillations are pronounced at the two ends of the bandwidth of the power
spectrum of the signal as the limits of the integration are curtailed from
the ideal $-\infty$ to $\infty$. As the noise is very high at low frequencies
the amplitude of the oscillations in $\Omega(f)$ is very less at low
frequencies.  It is easy to explain these oscillations
using the Cornu's spiral \cite{19}.  The Cornu's spiral does not get wound up
before the limits of the integration are reached. The thickness of the line
indicates the presence of sub structure in the power spectrum.
\section {Numerical results}
\label {numeric}
\subsection{correlations and shifts in parameters}
The signal waveform was obtained by numerically integrating equation
(\ref{cut}). We get time as a function of frequency which we then invert
to get frequency as a function of time. This is now used to generate the phase
and the amplitude of the signal. We take the initial phase and the time of
arrival of the signal to be  zero.
We now present the results of the numerical simulations. We have considered
black hole masses in the range 5 to 10 $M_\odot$.  The masses taken for
the smaller mass are 0.5, 1.0, and 1.4 $M_\odot$. The analysis has been
carried out for the  LIGO detector both in the initial and the advanced
stages. We retain $\xi$ as a parameter for the restricted post-Newtonian
waveform also defined by the equation (\ref{eqxi}) though this quantity does
not represent the coalescence time of the signal anymore. In general the amount
of time the signal lasts is less for the post-Newtonian signal as compared with
the Newtonian one which follows  from the fact that the quantity $F(x)$ in
equation (\ref{cut}) is less than one in the frequency range  considered.

In table \ref{tab1} we list the normalised correlations obtained for the
LIGO detector in the initial stage. The mass of the larger component of the
binary ($M_1$) increases from left to right along each row. The mass of
the other component ($M_2$) increases from top to bottom in each column.
The values of the masses are listed accordingly in the table. The correlations
show a very regular behavior  in the table. There are two factors
controlling the drop of the correlation:
\begin{enumerate}
\item increase of the magnitude of phase corrections with increase of total
mass of the binary system, and
\item	decrease of the integration time due to the increase of total mass
of the system.
\end{enumerate}
These two factors work against each other in producing the total amount
of phase error between the Newtonian filter and the restricted post-Newtonian
signal.
Thus when we increase $M_1$, the increase in the magnitude of the phase
corrections dominates over the loss in integration time and we
get  lower correlations when we go from left to right. Exactly the opposite
happens when we increase $M_2$ i.e. the effect of the decrease of integration
time dominates and the correlations increase.
 In table \ref{tab3} we list the correlations for
 the advanced LIGO. The same pattern is observed in this table too.
We observe that the correlations for the larger frequency range are smaller as
may be expected since the filter is more likely to go out of phase in  a
broader bandwidth. However, it should be emphasised that these correlations
are {\it normalised}. The correlation would be unity if the filter were
exactly matched to the signal. In absolute terms, if we consider a signal
with given parameters having the same amplitude then the correlation for
the advanced LIGO will be much larger than the initial LIGO since
the noise is less; first, we get a larger integration time and second, the
power spectral density is an order of magnitude less in the common bandwidth.
We find that for the parameters considered, the absolute values of the
correlations are larger by a factor of twenty for the advanced LIGO.

We next take up the issue of the shift in the parameters
 of the filters which produce maximum  correlations.
In table \ref {tab2} we list the shift in the parameters $\xi$ and
$t_a$ for the case of the initial LIGO detector. The phase parameter $\phi_0$
is an extremely sensitive parameter and its shifts are not regular.
The value of $\Delta\xi$ is always negative. This is because as mentioned
earlier a post-Newtonian signal will last for a smaller length of time as
compared to a Newtonian signal with the same values of $M_1$ and $M_2$. Also it
can be seen from equation (\ref{orbev}) and the definition of $\xi$ that the
first derivative of the frequency $\dot f$ is approximately
proportional to $\xi^{5/3}$. Therefore in order to obtain a higher value
of $\dot f$ the value of $\xi$ is reduced. Here again the factors, the
integration time and magnitude of phase corrections, compete against each other
in determining how  $\Delta\xi$ varies with an increase in either of the
masses.
 The value of $\Delta\xi$ decreases with increase in $M_1$ and increases with
increase in $M_2$.
Also $\Delta t_a$ is always negative. This parameter
tries to compensate for the reduction in the coalescence time by pushing the
filter
forward in time. As the table shows there is apparently a very strong
covariance
between these two parameters. The value of $\Delta t_a$ also decreases with
increase in $M_1$ and increases with increase in $M_2$. Typically for
$M_1 = 5 M_\odot$, and $M_2 = 1.4 M_\odot$ we get shifts of $\Delta\xi =
-1.32$ secs and $\Delta t_a = -1.218$ secs. This should be compared with
the coalescence time of the waveform which is about 9 secs.

 For the case of the advanced LIGO detector the magnitude of the shifts is
much more, but  the time for which the signal spends
in the frequency range 10 to 400 Hz is about a factor of fifty more than
that for the
initial LIGO  for similar masses. Here the same pattern
is observed in the variation of $\Delta\xi$ and $\Delta t_a$ as in the initial
LIGO.
The typical values of the shifts observed are $\Delta\xi = -2.7 $ secs and
 $\Delta t_a = -6.0912$ secs for $M_1 = 5 M_\odot$, and $M_2 = 1.4 M_\odot$.

	Simulations were also done for the band limited white noise with
the power spectral density having a constant value between 40 to 400 Hz.
 The results were  compared with those of  the initial stage of LIGO.
The effect of coloured noise of the type considered
here is to narrow band the signal. Thus the Newtonian filter has to match with
the signal over a smaller range of frequencies. However if the
narrow banding occurs at  higher frequencies, for the chirp,
 the magnitude of the phase
corrections is more. Thus in addition to changing the first derivative
of the frequency through the parameter $\xi$ the values of the higher
derivatives of the frequencies would also have to be changed to get
a good match.
Had the narrowbanding been at lower frequencies where the time derivatives
of the frequency are relatively less, the correlation would have been
much higher as the shifts in the Newtonian parmeters would have been
sufficient for the purpose.
In table \ref {tab5} we show the correlations obtained
for band limited white noise  and table \ref {tab6} shows the corresponding
 parameter shifts. We observe that in  the case of the initial LIGO
the correlations  obtained are less than  those for the white noise case
 for higher values of the total mass and vice-versa. This can also be seen
as an effect of narrow banding. The values of the
shifts in the parameters is also much smaller in the case of white noise as
a most of the  contribution  to the correlation comes from the lower
frequencies
where a small
shift in $\xi$ is sufficient for the filter to match well with the signal.

\subsection{Effect of discreteness of the bank of filters}
Till now we have considered our filter bank to have an infinite number of
filters i.e. we have allowed for a continuous variation  of $\xi$.
 However one is limited by the computing power available and one must confine
oneself to a finite number of filters. Thus in general the signal
will be unable to achieve its maximum correlation.
 Our aim is to estimate the drop in the correlation for a given computing
speed.
The maximum drop in the
correlation because of the finiteness of the filter bank will
have to be kept small. We consider a discrete set of
Newtonian filters corresponding to distinct values of the $\xi$ parameter. The
filter spacing in the $\xi$ parameter is taken to be constant ($\delta\xi_c$)
 across the entire range of values $\xi$ can take (see \cite{5}).

We first consider the initial LIGO and assume a
 1 Gigaflop machine on which we intend to do on-line search. The
maximum time the signal lasts is found to be 25 secs for the mass range
considered. However the data train needs to be padded with zeroes to four times
the original length which is   optimal for computational purposes (see
\cite{11}). This   will increase the length of the data train to 100 secs. We
allow for an overlap of 25 secs between consecutive data trains.Thus we have 75
secs in which to calculate $n_f$ correlations where $n_f$ is the number of
filters.  We sample the
waveform at 1000 Hz. Thus we get approximately $2^{17}$ data points per
data train. We have to perform one Fast Fourier transform (FFT) operation per
filter. The Fourier transforms will have already been calculated once and for
all for the filters in the bank and one inverse Fourier transform will have to
be performed to obtain the correlation as a function of the time lag $\Delta
t$.
The computation time will be mostly taken up by the FFTs
as each FFT involves  $3N\log(N)$ operations where $N$ is the number of points
in the data train. In this particular case therefore each Fourier transform
will
require about 6.4 Million Floating point Operations (MFO).  This has to be
compared with the number
of floating point operations which can be carried out over the period of 75
secs
which is $75\times 10^9$ MFOs. Thus the number of filters that can be
accommodated is about 11700 filters. We require two filters for the phase
for each value of $\xi$. As the maximum value
of the coalescence time $\xi$ is 25 secs for the range of masses considered,
we get a filter spacing in the $\xi$ parameter to be around 4.3 msecs. In the
case of the advanced LIGO this number is about 172 msecs.

Let $\Delta\bar{\xi} = \Delta\xi - \Delta\xi_m$ where $\Delta\xi_m$ is the
value of $\Delta\xi$ corresponding to the maximum correlation.
 Figure \ref{fig4} shows how the correlation for a given signal
normalized to its maximum value
varies with $\Delta\bar{\xi}$  along a line of curvature i.e.
along the curve parameterised by $\Delta\bar{\xi}$ along which  the drop in the
correlation is the slowest. In other words the figure shows the correlation
 maximized over $\Delta t_a$ and $\Delta\phi_0$
as  a function of $\Delta\bar{\xi}$.
 The curve has been plotted for the initial LIGO and for $M_1 = 5M_\odot$,
and $M_2 = 1.4M_\odot$. However the shape of this curve and the magnitudes in
the drop of the correlation  is insensitive
to the values of $M_1$ and $M_2$.
We observe that even for shifts of 100msecs the correlation does not
drop by more than $2\%$. For the case of the advanced LIGO this drop in the
correlation is even lower.  Thus the filter spacing which we have calculated is
sufficient for our purpose.

\section{conclusion}
\label{end}

	We have demonstrated here the possibility of using Newtonian filters
for detecting the presence of  a restricted post-Newtonian signal.
 Such a strategy would be very useful in providing a
preliminary on line analysis of the data train. The analysis which we
have carried out here is valid only for the point mass case where $\mu << M $
where $\mu$ is the reduced mass and $M$ the total mass. For the initial
 LIGO the correlation is 0.65 on an average and for the advanced
LIGO it is around 0.45. These are only the normalised correlations as
we have already stressed before.  The absolute values of the signal to noise
will be much higher (by a factor of about 20) for the advanced LIGO.
It must be noted that the  drop of the correlation will translate
into a loss in the event rate. The distance upto which we can detect
the binary will come down by a factor equal to the normalised correlation. This
means that for the initial LIGO the distance to which we can detect the
binary will be brought down by $35\%$ and for the advanced LIGO
it will be brought down by  $55\%$ from their respective maximum ranges.
In absolute
terms the advanced LIGO will still be able to look further than the initial
LIGO.

	The effect of the discreteness of the filter bank in producing a
further drop in the correlations was investigated. It was found that for
a one Gigaflop machine the drop in correlation due to the discreteness
was very small. With better and faster machines we can make the bank
of filters still more efficient.

If we consider higher derivatives of frequency $f$ say $\ddot f$ {\it etc.}
 as parameters \cite{6} we should get a better match,  but the computation is
very
likely to increase. It should be possible to construct filters which not only
enable us to save on the computation time but also span the set of signal
waveforms adequately. A deeper analysis of the signal waveforms is in order so
that efficient  techniques can be developed. This work is now in progress.

\acknowledgments
R.B. would like to thank Dr B.S. Sathyaprakash for many fruitful discussions
and
suggestions. R.B. is being supported by the Junior Research Fellowship of CSIR.

\begin{figure}
\caption{The impulse response of a typical Newtonian filter is shown.
The value of the coalescence time for the Newtonian signal corresponding
to the filter is 9.1703 secs. The other two parameters, the arrival time
and initial phase are set to zero.}
\label{fig1}
\end{figure}

\begin{figure}
\caption{The figure illustrates how well the post-Newtonian signal and the
optimally matched Newtonian filter correlate. The best matching occurs
where the signal strength is high i.e. near coalescence. The three figures
correspond to the frequency ranges 103.40 to 109.56 Hz, 149.24 to 168.90 Hz
and  234.856 to 400.0 Hz respectively. }
\label{fig2}
\end{figure}

\begin{figure}
\caption{The figure shows $\Omega(f)$ in the frequency range 40 to 400 Hz. The
thickness of the line is due to the substructure present in the fourier
transform of the signal. }
\label{fig3}
\end{figure}

\begin{figure}
\caption{ Variation of the normalised correlation with $\Delta\xi$ along
the line of curvature which is the curve along which the matching factor
falls least.}
\label{fig4}
\end{figure}
\newpage
\mediumtext
\begin{table}
\caption{This table displays the correlations for the
initial LIGO detector for a wide range of masses in
units of solar masses. The black hole mass varies from 5 to 10 $M_\odot$
 and the other mass takes the values 0.5, 1.0 and 1.4 $M_\odot$. \label{tab1}}
\begin{tabular}{ccccccc}
&5.0$M_\odot$&6.0$M_\odot$&7.0$M_\odot$&8.0$M_\odot$&9.0$M_\odot$&10.0$
M_\odot$ \\
\tableline
0.5$M_\odot$&0.6403&0.6204&0.5979&0.5830&0.5762&.5608\\
&&&&&&\\
1.0$M_\odot$&0.7182&0.7045&0.6999&0.6838&0.6711&0.6411\\
&&&&&&\\
1.4$M_\odot$&0.7474&0.7434&0.7309&0.7269&0.7190&0.6988\\
&&&&&&\\
\end{tabular}
\end{table}

\begin{table}
\caption{Shown below are the shifts $\Delta\xi$ and $\Delta t_a$ in secs
($\Delta\xi$ above and $\Delta t_a$ below) in the case
of the initial LIGO detector. \label{tab2}}
\begin{tabular}{ccccccc}
&5.0$M_\odot$&6.0$M_\odot$&7.0$M_\odot$&8.0$M_\odot$&9.0$M_\odot$&10.0$
M_\odot$ \\
\tableline
&&&&&&\\
&$-$3.300&$-$3.800&$-$4.200&$-$4.650&$-$5.100&$-$5.4\\
0.5$M_\odot$&$-$3.160&$-$3.540&$-$3.835&$-$4.180&$-$4.541&$-$4.753\\
&&&&&&\\
&$-$1.660&$-$1.875&$-$1.980&$-$2.310&$-$2.460&$-$2.56\\
1.0$M_\odot$&$-$1.552&$-$1.708&$-$1.760&$-$2.04&$-$2.144&$-$2.191\\
&&&&&&\\
&$-$1.320&$-$1.475&$-$1.560&$-$1.600&$-$1.640&$-$1.755\\
1.4$M_\odot$&$-$1.218&$-$1.331&$-$1.379&$-$1.385&$-$1.396&$-$1.478\\
&&&&&&\\
\end{tabular}
\end{table}

\begin{table}
\caption{This table displays the correlations for the
advanced LIGO detector. The value of masses is the same as in the
previous tables \label{tab3}}
\begin{tabular}{ccccccc}
&5.0$M_\odot$&6.0$M_\odot$&7.0$M_\odot$&8.0$M_\odot$&9.0$M_\odot$&10.0$
M_\odot$ \\
\tableline
0.5$M_\odot$&0.4420&0.4230&0.4078&0.3963&0.3866&0.3781\\
&&&&&&\\
1.0$M_\odot$&0.5118&0.4956&0.4812&0.4702&0.4607&0.4526\\
&&&&&&\\
1.4$M_\odot$&0.5420&0.5276&0.5157&0.5055&0.4978&0.4891\\
&&&&&&\\
\end{tabular}
\end{table}

\begin{table}
\caption{Shown below are the shifts $\Delta\xi$ and $\Delta t_a$ in secs
($\Delta\xi$ above and $\Delta t_a$ below) in the case
of the advanced LIGO detector.\label{tab4}}
\begin{tabular}{ccccccc}
&5.0$M_\odot$&6.0$M_\odot$&7.0$M_\odot$&8.0$M_\odot$&9.0$M_\odot$&10.0$
M_\odot$ \\
\tableline
&&&&&&\\
&$$-$$3.000&$-$7.700&$-$11.495&$-$15.90&$-$19.20&$-$22.80\\
0.5$M_\odot$&$-$12.09&$-$15.64&$-$18.44&$-$21.94&$-$24.42&$-$27.263\\
&&&&&&\\
&$-$2.750&$-$4.750&$-$7.750&$-$9.250&$-$11.50&$-$13.00\\
1.0$M_\odot$&$-$7.421&$-$8.774&$-$11.22&$-$12.24&$-$14.05&$-$15.15\\
&&&&&&\\
&2.700&$-$4.440&$-$5.500&$-$7.050&$-$8.85&$-$9.450\\
1.4$M_\odot$&$-$6.091&$-$7.328&$-$7.971&$-$9.156&$-$10.63&$-$10.94\\
&&&&&&\\
\end{tabular}
\end{table}

\begin{table}
\caption{This table displays the  correlations for the case of band limited
white noise. The bandwidth ranges from 40 to 400 Hz.
 The masses are given in units of solar masses. The value of the masses
is same as in the previous tables. \label{tab5}}
\begin{tabular}{ccccccc}
&5.0$M_\odot$&6.0$M_\odot$&7.0$M_\odot$&8.0$M_\odot$&9.0$M_\odot$&10.0$
M_\odot$ \\
\tableline
0.5$M_\odot$&0.6363&0.6182&0.6043&0.5933&0.5838&.5765\\
&&&&&&\\
1.0$M_\odot$&0.6977&0.6850&0.6753&0.6659&0.6576&0.6517\\
&&&&&&\\
1.4$M_\odot$&0.7238&0.7135&0.7054&0.6969&0.6912&0.6874\\
&&&&&&\\
\end{tabular}
\end{table}

\begin{table}
\caption{Shown below are the shifts $\Delta\xi$ and $\Delta t_a$ in secs
($\Delta\xi$ above and $\Delta t_a$ below) in the case of band limited
white noise.\label{tab6}}
\begin{tabular}{ccccccc}
&5.0$M_\odot$&6.0$M_\odot$&7.0$M_\odot$&8.0$M_\odot$&9.0$M_\odot$&10.0$
M_\odot$ \\
\tableline
&&&&&&\\
&$-$.0385&$-$.135&$-$.210&$-$.288&$-$.370&$-$.43\\
0.5$M_\odot$&$-$.062&$-$.069&$-$.068&$-$.073&$-$.082&$-$.082\\
&&&&&&\\
&$-$.0735&$-$.125&$-$.165&$-$.205&$-$.240&$-$.280\\
1.0$M_\odot$&$-$.051&$-$.055&$-$.055&$-$.057&$-$.058&$-$.063\\
&&&&&&\\
&$-$.084&$-$.12&$-$.15&$-$.184&$-$.215&$-$.235\\
1.4$M_\odot$&$-$.047&$-$.049&$-$.050&$-$.055&$-$.059&$-$.057\\
&&&&&&\\
\end{tabular}
\end{table}
\end{document}